\documentclass[aps,prl,twocolumn,superscriptaddress,showpacs]{revtex4-1}
\usepackage{amsmath}
\usepackage{graphicx}
\usepackage{epsfig}
\usepackage{bm}
\usepackage{hyperref}

\def\be{\begin{equation}}
\def\ee{\end{equation}}
\def\bea{\begin{eqnarray}}
\def\eea{\end{eqnarray}}
\def\bean{\begin{mathletters}\begin{eqnarray}}
\def\eean{\end{eqnarray}\end{mathletters}}

\begin{document}


\title{An Equation of State of a Carbon-Fibre Epoxy Composite under Shock Loading}

\author{Alexander A. Lukyanov}
\affiliation{Abingdon Technology Centre, Schlumberger, Abingdon,
OX14 1UJ, UK}

\begin {abstract}
An anisotropic equation of state (EOS) is proposed for 
the accurate extrapolation of high-pressure shock Hugoniot (anisotropic and 
isotropic) states to other thermodynamic (anisotropic and isotropic) 
states for a shocked carbon-fibre epoxy composite (CFC) of any symmetry. 
The proposed EOS, using a generalised decomposition of a stress tensor [Int. 
J. Plasticity \textbf{24}, 140 (2008)], represents a mathematical and physical 
generalisation of the Mie-Gr\"{u}neisen EOS for isotropic material and reduces 
to this equation in the limit of isotropy. Although a linear relation between 
the generalised anisotropic bulk shock velocity $U^{A}_{s}$ and particle 
velocity $u_{p}$ was adequate in the through-thickness orientation, damage 
softening process produces discontinuities both in value and slope in the 
$U^{A}_{s}$-$u_{p}$ relation. Therefore, the two-wave structure (non-linear 
anisotropic and isotropic elastic waves) that accompanies damage softening 
process was proposed for describing CFC behaviour under shock loading. The 
linear relationship $U^{A}_{s}$-$u_{p}$ over the range of measurements 
corresponding to non-linear anisotropic elastic wave shows a value of 
$c^{A}_{0}$ (the intercept of the $U^{A}_{s}$-$u_{p}$ curve) that is in 
the range between first and second generalised anisotropic bulk speed of 
sound [Eur. Phys. J. B \textbf{64}, 159 (2008)]. An analytical calculation 
showed that Hugoniot Stress Levels (HELs) in different directions for a 
CFC composite subject to the two-wave structure (non-linear anisotropic 
elastic and isotropic elastic waves) agree with experimental measurements at 
low and at high shock intensities. The results are presented, discussed and 
future studies are outlined.
\end{abstract}
\date{\today}
\maketitle
\section{Introduction}
\label{intro}
Investigation of anisotropic composite materials (e.g., CFC
materials) behaviour has found significant interest in the
research community due to the widespread application of
anisotropic composite materials in aerospace and civil engineering
problems. Modern high-resolution methods for monitoring the stress
and particle velocity histories in shock waves and equipment have
been created
\cite{BarkerHollenbach,Kanel,Kaneletal,BourneStevens,Bourne}. A
common technique for the study of material behaviour under shock
loading is the planar plate impact test (one-dimensional shock
wave propagation). Shock wave experiments have frequently provided
the motivation for the construction of material constitutive
relations and have been the principal means for determining
material parameters for some of these relations
\cite{DavisonGraham,Bushmanetal,Meyers,Steinberg,KiselevLukyanov}.
For example, Dandekar {\it et al.} \cite{Dandekaretal}
investigated the equation of state of a glass fibre epoxy
composite, in terms of the shock stress, shock velocity $U_{s}$
(through-thickness orientation) and particle velocity $u_{p}$
(i.e. the velocity of material flow behind the shock front). Their
results indicated that there was a linear relationship between
shock and particle velocity. This type of behaviour is typical of
a wide range of materials, including metals
\cite{Meyers,Steinberg} and some polymers, including epoxy resins
\cite{MunsonMay,Millettetal}, and composite \cite{Zhuketal},
including  carbon fibre epoxy composite \cite{Riedeletal} and
glass fibre epoxy composite \cite{Zaretskyetal}. A linear
$U_{s}$-$u_{p}$ relationship shows that in the through thickness
orientation, this class of composite displays fairly typical
experimental data. However, in spite of a perfectly adequate
general understanding, experimental methodology, and theory,
material models do not agree in detail, especially for anisotropic
composite materials. For many years, it has been assumed that the
response of composite materials to shock loading is isotropic
\cite{Chenetal,Hayhurstetal,Andersonetal1}, and only recently has
anisotropy in the shock response of anisotropic materials (e.g.,
composite materials) attracted the attention of researchers
\cite{Andersonetal2,Bordzilovskyetal,Hereiletal}, 
\cite{MillettLukyanov,Lukyanov2,Lukyanov3,Lukyanov4}.

In this paper, a macroscopic continuum framework is chosen to
modify the existing methodology for accurate extrapolation of
high-pressure shock Hugoniot states to other thermodynamic states
for a shocked isotropic continuum. The composite materials
response under shock loading leads to a nonlinear behavior (i.e.
large compressions), therefore, an equation of state (EOS) is
required \cite{Andersonetal1,Lukyanov2,Lukyanov3,Lukyanov4}. To
address this issue, an anisotropic equation of state similar to
\cite{Lukyanov2,Lukyanov3,Lukyanov4} is proposed for accurate
extrapolation of high-pressure shock Hugoniot (anisotropic and 
isotropic) states to other thermodynamic (anisotropic and 
isotropic) states for a shocked carbon-fibre epoxy composite
of any symmetry.
\section{An anisotropic Equation of State}
\label{sec:2}
The proposed equation of state, using the generalised decomposition of
a stress tensor \cite{Lukyanov2,Lukyanov3,Lukyanov4,Lukyanov1,Lukyanov5}, 
represents a mathematical and physical generalisation of the 
Mie-Gr\"{u}neisen equation of state for isotropic material and 
reduces to this equation in the limit of isotropy. The generalised 
decomposition of the stress tensor is summarised below. The generalised 
decomposition framework will provide a useful point of construction of
an anisotropic equation of state.
\subsection{Generalised decomposition of the stress tensor:
$\alpha$-$\beta$ decomposition}
\label{sec:3}
The definition of pressure in the case of an anisotropic solids
should be the result of stating that the "pressure" term should
only produce a change of scale, i.e. isotropic state of strain
\cite{Lukyanov2,Lukyanov3,Lukyanov4,Lukyanov1,Lukyanov5}. The 
generalised decomposition of the stress tensor $\sigma_{ij}$ is 
defined as:
\begin{equation}
\label{Luk3_1} \displaystyle \tilde{{\rm P}}:\tilde{S}= 0 \quad
\mbox{or}\quad\alpha_{ij}\tilde{S}_{ij}=0, \quad \sigma_{ij}=-
p^{\ast} \alpha_{ij} + \tilde{S}_{ij},
\end{equation}
where $\tilde{{\rm P}}=p^{\ast}\alpha_{ij}$ is the generalised
spherical part of the stress tensor, $\tilde{S}=\tilde{S}_{ij}$ is
the generalised deviatoric stress tensor, $p^{\ast}$ is the total
generalised "pressure" and $\alpha_{ij}$ is the first
generalisation of the Kronecker delta symbol. The constructive
definition of the tensor $\alpha_{ij}$ is based on the fact that
the stress tensor $p\alpha_{ij}$ is induced in the anisotropic medium
by the applied isotropic strain tensor
$\displaystyle\frac{\varepsilon_{v}}{3}\delta_{ij}$, i.e.
\begin{equation}
\label{Luk3_1_1} \displaystyle
p\alpha_{ij}=-\frac{\varepsilon_{v}}{3}\textbf{C}_{ijkl}\delta_{kl},
\quad p=-K_{C}\varepsilon_{v},
\end{equation}
where $p$ is the pressure, $\varepsilon_{v}$ is the volumetric
strain, $\delta_{ij}$ is the Kronecker delta symbol (unit
tensor), $\textbf{C}_{ijkl}$ is the elastic stiffness matrix and
$K_{C}$ is the first generalised bulk modulus. The expressions for
the $\alpha_{ij}$ components and $K_{C}$ are presented below. Throughout, 
the contraction by repeating indexes is assumed. 
Using (\ref{Luk3_1}), the following expression for total 
generalised "pressure" $p^{\ast}$ can be obtained:
\begin{equation}
\label{Luk3_2} \displaystyle p^{\ast}=
-\frac{{\sigma_{ij}\alpha_{ij}}}{{\alpha_{kl}\alpha_{kl}}}=
-\frac{{1}}{{\left\|{\alpha}\right\|^{2}}}\sigma_{ij}\alpha _{ij},
\end{equation}
where $\left\|{\alpha}\right\|^{2}=\alpha_{ij}\alpha_{ij}
=\alpha_{11}^{2}+\alpha_{22}^{2}+\alpha_{33}^{2}$. Finally,
the expression for the generalised deviatoric part
of the stress tensor can be rewritten in the following form:
\begin{equation}
\label{Luk3_3}\displaystyle
\tilde{S}_{ij}=\sigma_{ij}-\alpha_{ij}\cdot
\frac{{1}}{{\left\|{\alpha}\right\|^{2}}}\sigma_{kl}\alpha_{kl} .
\end{equation}
For anisotropic materials, the total generalised "pressure"
$p^{*}$ has been expressed 
\cite{Lukyanov2,Lukyanov3,Lukyanov4,Lukyanov1,Lukyanov5} as:
\begin{equation}
\label{Luk3_4}\displaystyle p^{*}=p+p^{\tilde{S}}, \
p = -\frac{\beta_{ij}\sigma_{ij}}{\beta_{ij}\alpha_{ij}},
\ p^{\tilde{S}}=\ \frac{{\beta_{ij}\tilde
{S}_{ij}}}{{\beta_{kl}\alpha_{kl}}},
\end{equation}
where $p$ is the pressure related to the volumetric deformation
(\ref{Luk3_1_1}), $p^{\tilde{S}}$ is the pressure related to the
generalised deviatoric stress and $\beta_{ij}$ is the second
generalisation of the Kronecker delta symbol. The constructive
definition of the tensor $\beta_{ij}$ is based on the fact that
stress tensor $p\beta_{ij}$ is independent of the stress tensor
$\textbf{C}_{ijkl}e_{kl}$, i.e. their contraction product is zero
for any deviatoric strain tensor $e_{kl}$, where
$\textbf{C}_{ijkl}$ is the elastic stiffness matrix. The following
relation describes the functional definition of the second order
material tensor $\beta_{ij}$:
\begin{equation}
\label{Luk3_4_1}\displaystyle \beta_{ij}\textbf{C}_{ijkl}
=3K_{S}\delta_{kl},
\end{equation}
where $K_{S}$ represents the second generalised bulk modulus. The
solution of equations (\ref{Luk3_4_1}) in terms of the $\beta_{ij}$
components and an expression for $K_{S}$ are presented below.
Equations (\ref{Luk3_4}) define the correct generalised "pressure"
for the elastic regime. Note that the generalised decomposition of
the stress tensor can be applied for all anisotropic solids of any
symmetry and represents a mathematically consistent generalisation
of the conventional isotropic case. The procedure of construction
for the tensor $\alpha_{kl}$ has been defined in
\cite{Lukyanov2,Lukyanov3,Lukyanov4,Lukyanov1,Lukyanov5}. 
The elements of the tensor
$\alpha_{kl}$ are
\begin{equation}
\label{Luk3_5}
\begin{array}{*{20}c}
\displaystyle\alpha _{11}=\left(\sum^{3}_{k=1}C_{k1}\right)\cdot
3\bar {K}_{C}, \
\alpha _{22}=\left(\sum^{3}_{k=1}C_{k2}\right)\cdot 3\bar {K}_{C},
\hfill \\
\displaystyle\alpha _{33}=\left(\sum^{3}_{k=1}C_{k3}\right)\cdot
3\bar {K}_{C}, \
\alpha_{ij}\alpha_{ij}=3,\hfill
\end{array}
\end{equation}
\begin{equation}
\label{Luk3_6}
\begin{array}{*{20}c}
\displaystyle K_{C} =
\frac{1}{3\sqrt{3}}
\sqrt{\left(\sum^{3}_{k=1}C_{k1}\right)^{2}+
\left(\sum^{3}_{k=1}C_{k2}\right)^{2}+
\left(\sum^{3}_{k=1}C_{k3}\right)^{2}},\hfill \\
\displaystyle K_{C} = \frac{1}{9\bar{K}_{C}}, \hfill
\end{array}
\end{equation}
where $C_{ij}$ is the elastic stiffness matrix (written in
Voigt notation). The elements of the tensor
$\beta_{kl}$ are
\begin{equation}
\label{Luk3_7}
\begin{array}{*{20}c}
\displaystyle \beta_{11} =\left(\sum^{3}_{k=1} J_{k1}\right)\cdot 3K_{S}, \
\beta_{22} =\left(\sum^{3}_{k=1} J_{k2}\right)\cdot 3K_{S}, \hfill \\
\displaystyle \beta_{33} =\left(\sum^{3}_{k=1} J_{k3}\right)\cdot 3K_{S}, \
\beta_{ij}\beta_{ij}=3, \hfill
\end{array}
\end{equation}
\begin{equation}
\label{Luk3_8}
\displaystyle\frac{1}{K_{S}}=
\sqrt{3}\sqrt{\left(\sum^{3}_{k=1}J_{k1}\right)^{2}
+\left(\sum^{3}_{k=1}J_{k2}\right)^{2}+
\left(\sum^{3}_{k=1}J_{k3}\right)^{2}},
\end{equation}
where $J_{ij}$ are elements of the compliance matrix (written in Voigt
notation). In the limit of isotropy, the proposed generalisation
returns to the traditional classical case where tensors
$\alpha_{ij}$ and $\beta_{ij}$ equal $\delta_{ij}$ and equations
(\ref{Luk3_4}) take the form:
\begin{equation}
\label{Luk3_9}
\begin{array}{*{20}c}
\displaystyle
p^{\ast}=-\frac{{\sigma_{ij}\delta_{ij}}}{{\delta
_{kl}\delta_{kl}}}=-\frac{{1}}{{3}}\sigma_{kk}
\hfill \\
\displaystyle
p = -
\frac{{\beta_{ij}\sigma_{ij}}}{{\beta_{kl}\alpha_{kl}}}= -
\frac{{1}}{{3}}\sigma_{kk},
\hfill\\
\displaystyle
p^{\tilde{S}}=\ \frac{{\beta_{ij}\tilde
{S}_{ij}}}{{\beta_{kl}\alpha_{kl}}}=0,\hfill
\end{array}
\end{equation}
where $\displaystyle p=-\frac{\sigma_{kk}}{3}$ is the classical
hydrostatic pressure. Also, parameters $K_{C}$ and $K_{S}$ were
considered as the first and the second generalised bulk moduli. In
the limit of isotropy, they reduce to the well-know expression for
conventional bulk modulus.

\begin{figure}
\includegraphics[width=3.0in,height=2.5in]{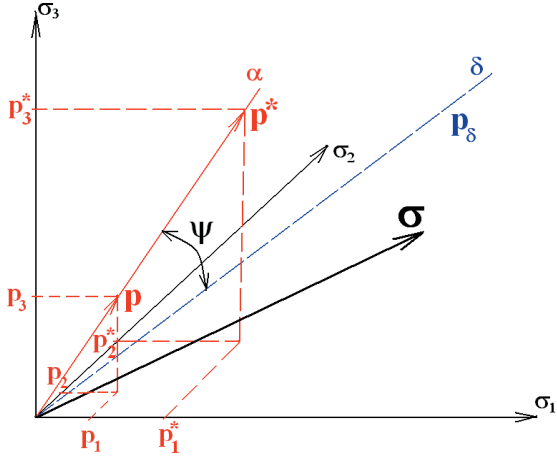}
\caption{Schematic representation of $\alpha$-decomposition of the
stress tensor (Haigh-Westergaard stress space).}
\label{fig:1}
\end{figure}

The geometrical representation of generalized decomposition of the
stress can be shown in the principal stress space
(Haigh-Westergaard stress space) for $\alpha$-decomposition of the
stress tensor and $\beta$-decomposition of the stress tensor in
the principal strain space. Figure \ref{fig:1} shows schematic
representation of $\alpha$-decomposition of the stress tensor,
where $\alpha$-direction is described by the tensor $\alpha_{ij}$
and $\delta$-direction is described by the Kronecker's delta
tensor $\delta_{ij}$. Therefore, $p_{\delta}$ describes the
hydrostatic stress (isotropic stress), $p^{\ast}\neq p_{\delta}$
describes the total generalized hydrostatic stress (or anisotropic
total generalized hydrostatic stress), and $p\neq p^{\ast}\neq
p_{\delta}$ is the generalized pressure related to an equation of
state (EOS). The angle between $\alpha$-direction and
$\delta$-direction is described by the variable $\psi$ which can
be obtained from $\displaystyle\cos\psi
=\frac{\alpha_{11}+\alpha_{22}+\alpha_{33}}{3}$. Similar
representation can be shown for $\beta$-decomposition of the
stress tensor (Haigh-Westergaard stress space).
%
%
\subsection{An anisotropic EOS}
\label{sec:4}
The equations (\ref{Luk3_4}) define the correct generalised
"pressure" for the elastic regime:
\begin{equation}
\label{Luk4_1}\displaystyle p^{*}=p+\frac{{\beta_{ij}\tilde
{S}_{ij}}}{{\beta_{kl}\alpha_{kl}} }, \
\sigma_{ij} = -p^{*}\alpha_{ij}+ \tilde{S}_{ij}.
\end{equation}
Further, to provide an appropriate description for general anisotropic 
materials behavior at high pressures, the pressure $p$ related to the 
volumetric deformations is described by the Mie-Gr\"{u}neisen equation 
of state pressure $p^{EOS}$:
\begin{equation}
\label{Luk4_1_1}
p=p^{EOS}=f\left({\rho ,e}\right)=P^{A}_{H}\cdot\left({1-\frac{{\Gamma
}}{{2}}\mu}\right)+\rho\Gamma e
\end{equation}
or
\begin{equation}
\label{Luk4_2}\displaystyle p^{EOS} =
\left\{{{\begin{array}{*{20}c} \displaystyle
{\frac{{\rho_{0}\left(c^{A}_{0}\right)^{2}\mu \left[ {1 + \left(
{1 -\frac{{\Gamma(\nu)}}{{2}}} \right)\mu -
\frac{{\Gamma(\nu)}}{{2}}\mu^{2}} \right]}}{{\left[ {1 - \left(
{S^{A}_{1} - 1} \right)\mu - S^{A}_{2} \frac{{\mu^{2}}}{{\mu + 1}}
- S^{A}_{3} \frac{{\mu^{3}}}{{\left( {\mu
+ 1} \right)^{2}}}} \right]^{2}}}} + \hfill \\
{+\left( {1+ \mu}  \right) \cdot \Gamma(\nu) \cdot E, \quad
\mu > 0;} \hfill \\
{\rho_{0}\left(c^{A}_{0}\right)^{2}\mu + \left( {1 + \mu}  \right) \cdot
\Gamma \cdot E,\quad \mu < 0;} \hfill \\
\end{array}} } \right.,
\end{equation}
\begin{equation}
\label{Luk4_3} \displaystyle\Gamma (\nu) = \frac{{\gamma_{0} + a\mu} }{{1
+ \mu} }, \ \displaystyle\mu = \frac{{\rho} }{{\rho_{0}}}-1,
\end{equation}
where $\gamma_{0}$ is the initial Gr\"{u}neisen gamma, $a$ is the
first order volume correction to $\gamma_{0}$, $P^{A}_{H}$ is the
anisotropic generalised bulk Hugoniot pressure, $E$ is the
internal energy per initial density
$\displaystyle\left(E=\frac{e}{\rho_{0}}\right)$, $\mu$ is the
relative change of volume, $\nu$ is the specific volume and
$S^{A}_{1}$, $S^{A}_{2}$, $S^{A}_{3}$ are the intercept of the
cubic $U^{A}_{s}$-$u_{p}$ curve \cite{Steinberg}:
\begin{equation}
\label{Luk4_4}\displaystyle U^{A}_{s} = c^{A}_{0} + S^{A}_{1}
u_{p} + S^{A}_{2} \left({\frac{{u_{p}} }{{U^{A}_{s}}}}\right)u_{p}
+ S^{A}_{3}\left({\frac{{u_{p}} }{{U^{A}_{s}}}}\right)^{2}u_{p} ,
\end{equation}
where $U^{A}_{s}$ is the generalised anisotropic bulk shock
velocity and $c^{A}_{0}$ is the velocity curve intercept. The
proposed generalised "pressure" also correctly describes the
medium behavior at small volumetric strains. To be consistent
with the definition of the isotropic bulk speed of sound, the
following definitions of the first $c_{I}$ and the second $c_{II}$
bulk speed of sound for anisotropic medium are assumed
\cite{Lukyanov3}:
\begin{equation}
\label{Luk4_5} \displaystyle c_{I}=\sqrt\frac{K_{C}}{\rho_{0}},
\ c_{II}=\sqrt\frac{K_{S}}{\rho_{0}},
\end{equation}
where the generalised bulk moduli $K_{C}$, $K_{S}$ are defined
according to (\ref{Luk3_6}) and (\ref{Luk3_8}) respectively.
Parameters $c^{A}_{0}\in [c_{II},c_{I}]$, $S^{A}_{1}$,
$S^{A}_{2}$, $S^{A}_{3}$, $\gamma_{0}$, $a$ represent material
properties which define its EOS (\ref{Luk4_2}).
\section{The behaviour of a Carbon-Fibre Epoxy Composite under shock loading}
\label{sec:5}
The purpose of this section is to display the accurate extrapolation of
experimental Hugoniot states \cite{MillettLukyanov} behind the 
shock wave (thermodynamic states) to high-pressure shock Hugoniot 
Stress Levels (HSLs) for a shocked carbon-fibre epoxy 
composite (initially anisotropic CFC) using the anisotropic EOS proposed 
above.
\subsection{Description of Experiment}
\label{sec:6}
\begin{figure}
\includegraphics[width=8.5cm]{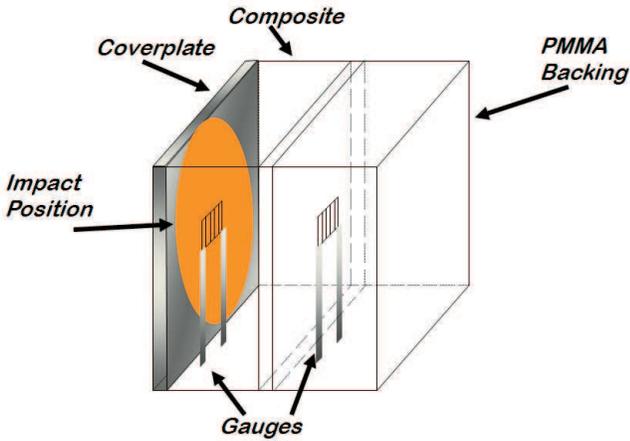}
\vspace{0.1cm}       
\caption{Target assembly.}
\label{fig:2}       
\end{figure}

The work discussed below concerns the shock response of a
carbon-fibre epoxy composite. This is done by the technique of
plate impact, whereby a flat plate of constant thickness and of
known material (for instance aluminium alloy, or copper) is
impacted onto a target plate made from the test material. The
flyer plates are launched using a $50 \ mm$ bore, $5 \ m$ long
single stage gas gun. The plate impact test was done at the Defence
Academy of the United Kingdom by Millett \textit{et al.}
\cite{MillettLukyanov} using samples of a carbon-fibre composite
(CFC) of thicknesses $3.8 \ mm$. The impact axises were 
(a) normal to the plane of the fibres and (b) parallel to the plane 
of the fibres. On impact, a planar shock front starts propagating 
into the target. The shock propagation in the target is monitored 
using manganin stress gauges, placed at different locations within 
the target assembly. A schematic of the target assembly and gauge 
placement is shown in Figure \ref{fig:2}. A manganin stress gauge 
was supported on the back of the specimen plate with a $12 \ mm$ 
block of polymethylmethacrylate (PMMA). Also, the gauge was 
backed into the PMMA by approximately $1.5 \ mm$ PMMA offset block 
to act as extra protection for the gauge. A second gauge (the $0 \ mm$ 
position) was supported on the front of the target assembly with a 
$1 \ mm$ plate of aluminium alloy 6082-T6. Shock stresses were 
induced with dural flyer plates impacting with the different velocities, 
using a single stage gas gun \cite{MillettLukyanov}.
%
%
The results from the stress gauges were converted to material
(Target) values $\sigma_{M}$, using the shock impedances of the
target $A_{T}$ and PMMA $A_{P}$, via the well-known relation
\cite{Meyers,Millettetal,MillettLukyanov}:
\begin{equation}
\label{Luk6_1}
\displaystyle\sigma_{M}=\frac{A_{T}+A_{P}}{2A_{P}}\sigma_{P},
\end{equation}
where $\sigma_{P}$ is the stress gauges values. Furthermore,
temporary spacing $(\delta t)$ between the tracers in 
combination with the specimen thickness  $(\delta w)$ were used
to obtain the shock velocities in the longitudinal (through thickness
orientation $U^{L}_{s}=\delta w/\delta t$ and along the fibre $0^{o}$ 
orientation $U^{F}_{s}=\delta w/\delta t$, taking into account both
the offset distance of the gauge within the PMMA and the known shock
response of PMMA \cite{BarkerHollenbach}, \cite{MillettLukyanov}.   
\subsection{Materials}
\label{sec:7} 
\begin{figure}
\includegraphics[width=8.5cm]{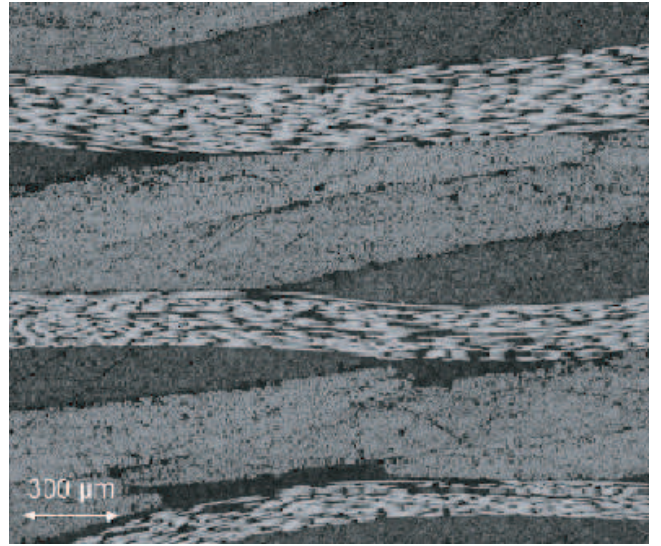}
\vspace{0.1cm}       
\caption{Microstructure of carbon-fibre epoxy composite.}
\label{fig:3}       
\end{figure}

The fibres in the carbon-fibre epoxy composite were
Hexcel 5HS in a woven lay up of orientation $[0/90,\pm 45]_{4}$.
The arial weight was $370 \ g/m^{2}$. The resin was an epoxy,
Hexcel RTM6. The composite was cured at $180^{o} \ C$ under a
pressure of $100 \ psi $ ($6.7 \ atm$) for 1 hour 40 minutes. The
microstructure is shown in Fig. (\ref{fig:3}). The longitudinal
sound speed in the through-thickness orientation $C_{L}$ was 
$3020 \ m/s$, and the ambient density $\rho$ was $1.50 \ g/cm^{3}$.
%
%
The equivalent material properties of the CFC composite plate were
chosen to match the layer macromechanical properties for the layup
$[0/90,\pm 45]_{4}$ only and the longitudinal sound speed in the
through-thickness orientation. A description of their numerical
values for the selected Carbon Fibre Composite (CFC) is shown in Table
\ref{tab:1}.
\begin{table}
\caption{Material properties of undamaged (CFC) Carbon Fibre
Composite (11-direction corresponds to the fill
direction, 22-direction to the warp direction
and 33-direction the through-thickness orientation).}
\label{tab:1}       
\begin{tabular}{lll}
\hline\noalign{\smallskip}
Parameter & Description & CFC  \\
\noalign{\smallskip}\hline\noalign{\smallskip}
$\rho^{A}_{0} \ \left[kg/m^{3}\right]$  & Density& 1500.0 \\
$E_{1}\left[GPa\right]$    & Elastic modulus (22 direction)& 68.467 \\
$E_{2}\left[GPa\right]$    & Elastic modulus (11 direction)& 66.537 \\
$E_{3}\left[GPa\right]$    & Elastic modulus (33 direction)& 13.678 \\
$K_{C}\left[GPa\right]$    & Firtst generalised bulk modulus& 19.436 \\
$K_{S}\left[GPa\right]$    & Second generalised bulk modulus& 7.6902 \\
$C_{L}\ \left[ m/s\right]$ & Sound speed (11 direction) & 6762.0 \\
$C_{L}\ \left[ m/s\right]$ & Sound speed (22 direction) & 6666.0 \\
$C_{L}\ \left[ m/s\right]$ & Sound speed (33 direction) & 3020.0 \\
$\nu_{21}$ &  Poisson's ratio & 0.0400  \\
$\nu_{31}$ &  Poisson's ratio & 0.0045  \\
$\nu_{32}$ &  Poisson's ratio & 0.0044  \\
$\alpha^{A}_{11}$ & Tensor $\alpha$ (22 direction) & 1.2290   \\
$\alpha^{A}_{22}$ & Tensor $\alpha$ (11 direction) & 1.1956   \\
$\alpha^{A}_{33}$ & Tensor $\alpha$ (33 direction) & 0.2454   \\
$\beta^{A}_{11}$  & Tensor $\beta$ (22 direction)  & 0.3155   \\
$\beta^{A}_{22}$  & Tensor $\beta$ (11 direction)  & 0.3254   \\
$\beta^{A}_{33}$  & Tensor $\beta$ (33 direction)  & 1.6717   \\
\noalign{\smallskip}\hline
\end{tabular}
\end{table}
The acoustic longitudinal speed of sound in the through-thickness 
orientation is:
\begin{equation}
\label{Luk7_1} \displaystyle
C_{L}=\sqrt{\left(\frac{1}{\rho}\frac{\partial \sigma_{L}}{\partial
\varepsilon_{L}}\right)_{\rho=\rho_{0}}},
\end{equation}
where $\sigma_{L}$ is the longitudinal stress, $\varepsilon_{L}$
is the longitudinal strain. Using the generalised Hook law for
orthotropic materials and an uniaxial acoustic strain state
(through thickness orientation), i.e.
$\varepsilon_{L}=\varepsilon_{33}=\varepsilon_{zz}$
($\varepsilon_{ij}\neq 0 \ i,j=3$ only), the longitudinal speed of
sound (through thickness orientation) (\ref{Luk7_1}) takes the form:
\begin{equation}
\label{Luk7_2} \displaystyle
C_{L}=
\sqrt{
\frac{1}{\rho_{0}}
\left[\frac{\left(1-\nu_{12}\nu_{21}\right)E_{3}}
{\Delta}
\right]},
\end{equation}
where
$\Delta=\left(1-\nu_{12}\nu_{21}-\nu_{23}\nu_{32}-\nu_{31}\nu_{13}
-2\nu_{21}\nu_{32}\nu_{13}\right)$; $E_{1}$, $E_{2}$, $E_{3}$ are
Young's moduli and $\nu_{ij}$ are Poisson ratios. The measured
longitudinal speed of sound was $C_{L}=3020\pm 30 \ m/s$ for the
initial material density $\rho_{0}=1500\pm 10 \ kg/m^{3}$
\cite{MillettLukyanov}. Using (\ref{Luk7_2}) and elastic material
properties in Table \ref{tab:1}, the longitudinal speed of sound
is calculated to be $C_{L}= 3020 \ m/s$, which is in exact
agreement with the measured value of $3020\pm 30 \ m/s$. Note that
we have a six elastic unknown material properties listed in
Table \ref{tab:1}, which directly affect shock waves propagation 
under uniaxial strain state. However, only one longitudinal sound 
speed in the through-thickness orientation, depending on four unknown 
elastic properties, was experimentally measured \cite{MillettLukyanov}. 
Hence, the equivalent material properties given in Table \ref{tab:1} 
have not been finalised yet for the selected CFC material.
\subsection{Results and discussion}
\label{sec:8}
Initially, the equations of state, in terms of the shock and particle 
velocities, are examined. Although a linear relationship in the 
$U^{A,L,F}_{s}$-$u_{p}$ plane is adequate for simple materials and 
some anisotropic metals \cite{Lukyanov2,Lukyanov3,Lukyanov4,Lukyanov6}, phase 
changes under shock loading (e.g., damage softening) produce discontinuities 
both in value and slope in the $U^{A,L,F}_{s}$-$u_{p}$ relation, where 
$U^{A,L,F}_{s}$ (i.e., $U^{A}_{s}$, $U^{L}_{s}$, $U^{F}_{s}$) 
represent the generalised shock velocities in the directions of anisotropic bulk,
longitudinal through-thickness, and along the fibre $0^{o}$ orientation, respectively. 
These discontinuities are usually caused by the two-wave structure that accompanies 
most phase changes, as well as damage softening. It was pointed out by Bethe \cite{Bethe} 
that, for stable shock waves, the shock velocity must increase with pressure. This 
means that if the shock velocity should decrease with pressure, then the shock 
front would break up into two or more waves, or possibly one wave with a 
continuously smeared front.

\begin{figure}
\includegraphics[width=8.5cm]{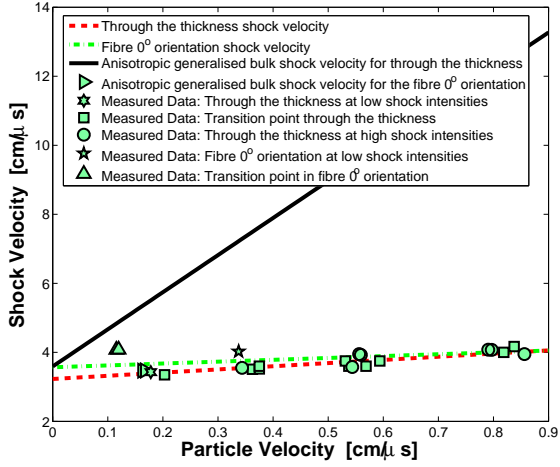}
\vspace{0.1cm}       
\caption{Anisotropic generalised bulk $U^{A}_{s}$ 
shock velocity, longitudinal (through the thickness) $U^{L}_{s}$ 
shock velocity and longitudinal (fibre $0^{o}$ orientation) $U^{F}_{s}$ 
shock velocity for a single wave structure of carbon fibre-epoxy 
composite under shock loading, where $U^{L}_{s}$ is defined 
by (\ref{Luk_Shock_Waves_Ani1}).} 
\label{fig:3a}       
\end{figure}

\begin{figure}
\includegraphics[width=8.5cm]{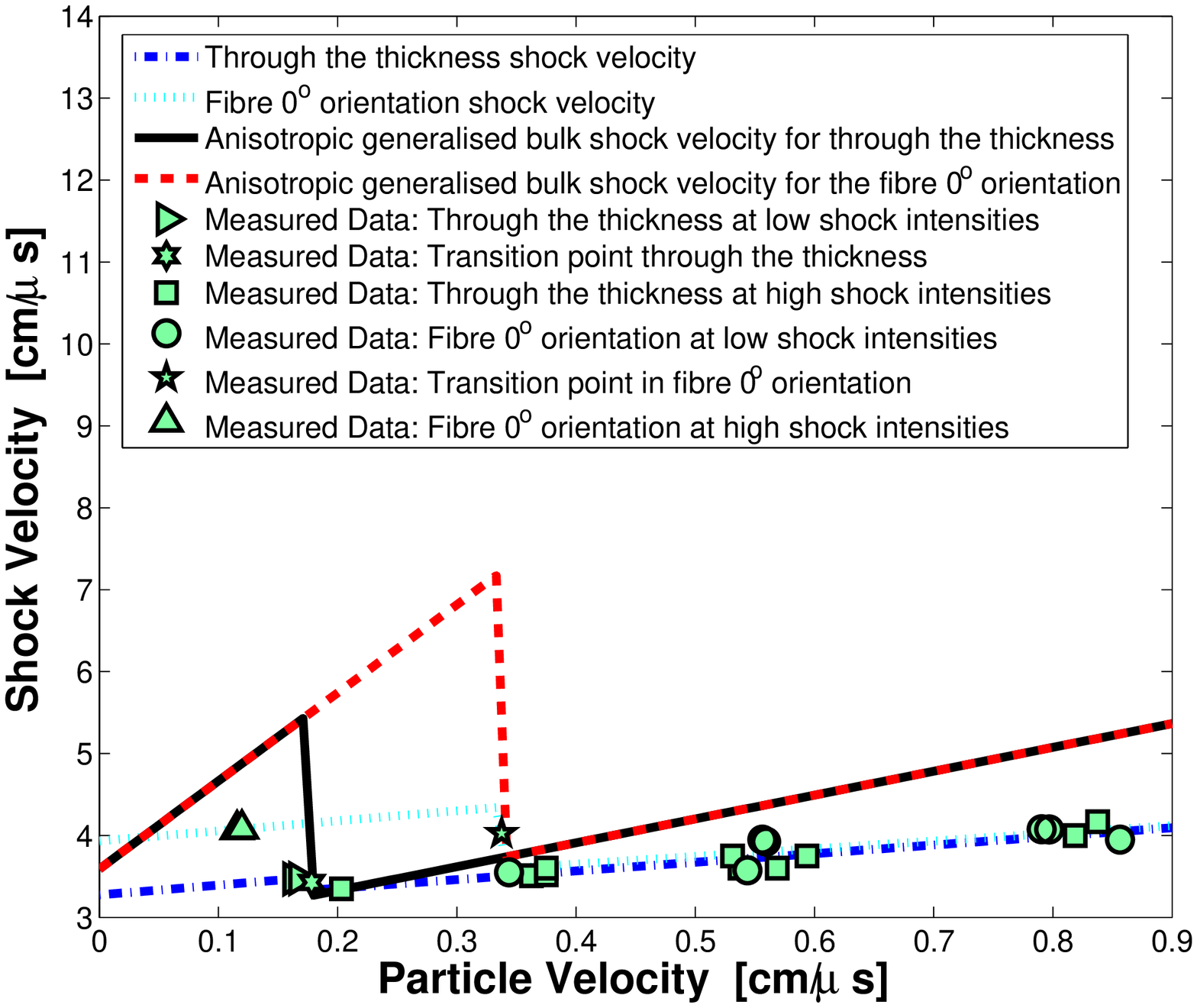}
\vspace{0.1cm}       
\caption{Anisotropic generalised bulk $U^{A}_{s}$ 
shock velocity, longitudinal (through the thickness) $U^{L}_{s}$ 
shock velocity and longitudinal (fibre $0^{o}$ orientation) $U^{F}_{s}$ 
shock velocity for a two-wave structure of carbon fibre-epoxy composite 
under shock loading, where $U^{L}_{s}$ is defined by 
(\ref{Luk_Shock_Waves_Ani2}).} 
\label{fig:4a}       
\end{figure}

Note that the experimental data for the through thickness
orientation (see, Fig. (\ref{fig:3a})) can be fitted by a 
linear relation, and there is no explicit evidence of the shock front breaking up;
however, the analysis of the experimental data for selected CFC material 
\cite{MillettLukyanov} (see, Fig. (\ref{fig:3a})) shows that the shock 
velocity along the fibre $0^{o}$ orientation decreases with pressure -- therefore, 
a two-wave structure is proposed for describing the experimental data. Additionally, 
a comparison of the equations of state in terms of Hugoniot Stress Levels (HSLs) 
for the single wave and two-wave structures is performed, in order to demonstrate the 
accuracy of the two-wave structure methodology. As a result, the experimental data 
\cite{MillettLukyanov} for longitudinal (through thickness) orientation 
and along the fibre $0^{o}$ orientation shock wave propagation in the 
selected CFC material has been fitted by straight lines (for single wave and 
two-wave structures), that is, using relationships for $U^{L}_{s}$-$u_{p}$ 
and $U^{F}_{s}$-$u_{p}$ that are linear (see, Figs. (\ref{fig:3a}), (\ref{fig:4a})).\\
For a single wave structure:
\begin{equation}
\label{Luk_Shock_Waves_1}
\begin{array}{*{20}c}
\displaystyle
a) \ U^{L}_{s}=c^{L}_{0}+S^{L}_{1}u_{p}, 
\ c^{L}_{0}, \ S^{L}_{1}>0, \ \forall \ u_{p},
\hfill \\
\displaystyle
b) \ U^{F}_{s}=c^{F}_{0}+S^{F}_{1}u_{p}, 
\ c^{F}_{0}, \ S^{F}_{1}>0, \ \forall \ u_{p},
\end{array}
\end{equation}
and, for a two-wave structure:
\begin{equation}
\label{Luk_Shock_Waves_2}
\begin{array}{*{20}c}
\displaystyle
c) \ U^{L}_{s}=\left\{
\begin{array}{*{20}c}
\displaystyle
c^{AL}_{0}+S^{AL}_{1}u_{p}, c^{AL}_{0}, S^{AL}_{1}>0, \ u_{p}\leq u^{*}_{l},
\hfill\\
c^{IL}_{0}+S^{IL}_{1}u_{p}, c^{IL}_{0}, S^{IL}_{1}>0, \ \ \ u_{p} > u^{*}_{l},
\end{array}\right.
\hfill\\
\displaystyle
d) \ U^{F}_{s}=\left\{
\begin{array}{*{20}c}
\displaystyle
c^{AF}_{0}+S^{AF}_{1}u_{p}, c^{AF}_{0}, S^{AF}_{1}>0, u_{p}\leq u^{*}_{f},
\hfill\\
c^{IF}_{0}+S^{IF}_{1}u_{p}, c^{IF}_{0}, S^{F}_{1}>0, \ \ \ u_{p} > u^{*}_{f},
\end{array}\right.
\end{array}
\end{equation}
where $u^{*}_{l}$, $u^{*}_{f}$ are the transition particle velocities oriented 
through the thickness and along the fibre $0^{o}$, respectively. 
Note that, as the severity of the shock increases, the Hugoniot Stress Levels 
(HSLs) of the two orientations converge. This fact demonstrates that the selected 
CFC material shows isotropic behaviour at high shock intensities, and can be 
described as an isotropic mixture of epoxy bunder and fractured fibres. Hence, 
$u^{*}_{l}$ and $u^{*}_{f}$ can be treated as transition particle velocities 
from the structured anisotropic material to the unstructured isotropic material. 
From the experimental data \cite{MillettLukyanov}, the following data is 
defined: $u^{*}_{l}=179.5\ m/s$ and $u^{*}_{f}=333.0\ m/s$ (see Table 
\ref{tab:3}). Note that, in the limit as $u^{*}_{p}, \ u^{*}_{f}\rightarrow\infty$, a 
two-wave structure methodology reduces to a single wave structure methodology. 
This property will be used to obtain EOS data for a single wave structure 
using the equations that apply to a two-wave structure. Therefore, the interpolation 
algorithm and equations below have been written for the two-wave structure only. 
Also, there is a degree of scatter within the experimental data; however, no 
obvious variation between shock velocities $U^{L}_{s}$, $U^{F}_{s}$ and 
specimen thickness has been observed \cite{MillettLukyanov}. Traditionally, 
the longitudinal (through thickness) orientation is tabulated better. Therefore, 
this direction was chosen for an accurate extrapolation of experimental HSLs 
to EOS data. 

Using a relation $U^{L}_{s}$-$u_{p}$ and the Rankine-Hugoniot 
equation expressing the conservation law of mass for a coordinate system 
in which the material in front of the shock wave is at rest, the
compressibility factor $\displaystyle\left(\frac{\rho^{*}_{0}}{\rho}\right)$ 
is calculated using,
\begin{equation}
\label{Luk7_3} \displaystyle
\rho(U^{L}_{s}-u_{p})=\rho^{*}_{0}U^{L}_{s}, \
\frac{\rho^{*}_{0}}{\rho}=\frac{U^{L}_{s}-u_{p}}{U^{L}_{s}}.
\end{equation}
Shock wave loading deals with large finite strains. Hence, the
Hencky strain (through thickness orientation), $\varepsilon^{H}$,
can be used to describe a uniaxial strain state for a
compression under shock loading \cite{Poirier}:
\begin{equation}
\label{Luk7_4}
\begin{array}{*{20}c}
\displaystyle
\varepsilon^{H}_{33}=\ln\left(\frac{\rho^{*}_{0}}{\rho}\right)=
\ln\left(\frac{U^{L}_{s}-u_{p}}{U^{L}_{s}}\right),
\hfill \hfill \hfill \hfill \\
\displaystyle
\varepsilon^{H}=\varepsilon^{H}_{33}, \ \varepsilon^{H}_{11}
=\varepsilon^{H}_{22}=0,
\ \mu=1-\frac{\rho_{0}}{\rho},
\hfill \hfill \hfill \hfill \\
\displaystyle
d^{H}_{33}=\frac{2}{3}\varepsilon^{H}, \
d^{H}_{11}=d^{H}_{22}=-\frac{1}{3}\varepsilon^{H},
\ d^{H}_{ij}\delta_{ij}=0,
\hfill \hfill \hfill\
\end{array}
\end{equation}
where $d^{H}_{ij}$ is the deviator of the Hencky strain tensor.
The experimental study of a carbon-fibre epoxy composite, shocking
along the through thickness orientation axis, showed no evidence of
an inelastic deformation. Therefore, elastic constitutive relations 
(before and after the transition zone) are considered here for 
approximating Hugoniot Stress Levels (HSLs) behind the shock wave:
\begin{equation}
\label{Luk7_5}\displaystyle
e=\frac{1}{2\rho^{*}_{0}}\textbf{C}^{*}_{ijkl}\varepsilon^{H}_{ij}
\varepsilon^{H}_{kl}, \ \sigma^{H}_{ij}=\rho\frac{\partial
e}{\partial \varepsilon^{H}_{ij}}, \
e=\frac{1}{2\rho}\sigma^{H}_{ij}\varepsilon^{H}_{kl},
\end{equation}
\begin{equation}
\label{Luk7_5_1}\displaystyle
\rho^{*}_{0},\textbf{C}^{*}_{ijkl},\alpha^{*}_{ij},\beta^{*}_{ij}
=\left\{
\begin{array}{*{20}c}
\displaystyle
\rho^{A}_{0},\textbf{C}^{A}_{ijkl},\alpha^{A}_{ij},\beta^{A}_{ij}, \ u_{p}\leq u^{*}_{p},
\hfill\\
\rho^{I}_{0},\textbf{C}^{I}_{ijkl},\alpha^{I}_{ij},\beta^{I}_{ij}, \ u_{p} > u^{*}_{p},
\end{array}\right.
\end{equation}
where $\sigma^{H}_{ij}$ is the Hencky stress tensor corresponding
to the Hencky strain tensor, $\rho^{*}_{0}$ is the initial density,
$\textbf{C}^{*}_{ijkl}$ is the elastic stiffness matrix, 
$\alpha^{*}_{ij}$ and $\beta^{*}_{ij}$ are the first and second generalisations of the
Kronecker delta symbol, $\rho^{A}_{0}$ is the initial density of the anisotropic CFC
material, $\rho^{I}_{0}$ is the released density of the isotropic (damaged) CFC
material, $\textbf{C}^{A}_{ijkl}$ is the anisotropic elastic
stiffness matrix calculated using the properties presented in Table
\ref{tab:1}, $\textbf{C}^{I}_{ijkl}$ is the isotropic elastic
stiffness matrix calculated using the properties presented in Table
\ref{tab:2}, $\alpha^{A}_{ij}$ and $\beta^{A}_{ij}$ are the first and second 
generalisations of the Kronecker delta symbol for the anisotropic CFC material 
(see Table \ref{tab:1}), $\alpha^{I}_{ij}$ and $\beta^{I}_{ij}$ are the first and 
second generalisations of the Kronecker delta symbol for the isotropic CFC material
(see Table \ref{tab:2}), and $u^{*}_{p}$ is the transition particle velocity, the later 
dependent upon the orientation (hence, $u^{*}_{p}=u^{*}_{l}$ for through the 
thickness orientation and $u^{*}_{p}=u^{*}_{f}$ for along the fibre $0^{o}$ 
orientation). 
%
%
\begin{table}
\caption{Material properties of damaged Carbon Fibre
Composite (CFC).}
\label{tab:2}       
\begin{tabular}{lll}
\hline\noalign{\smallskip}
Parameter & Description & CFC  \\
\noalign{\smallskip}\hline\noalign{\smallskip}
$\rho^{I}_{0} \ \left[kg/m^{3}\right]$  & Density& 1400.0 \\
$\lambda \left[GPa\right]$ & Parameter Lame (isotropic case) & 10.434 \\
$\mu     \left[GPa\right]$ & Shear modulus (isotropic case) & 0.18  \\
$C_{B}\ \left[ m/s\right]$ & Isotropic bulk sound speed & 2745.6 \\
$C_{L}\ \left[ m/s\right]$ & Sound speed (isotropic case) & 2777.0 \\
$\alpha^{I}_{11},\alpha^{I}_{22},\alpha^{I}_{33}$ & Tensor $\alpha_{ij}$ (isotropic case) & 1.0   \\
$\beta^{I}_{11},\beta^{I}_{22},\beta^{I}_{33}$  & Tensor $\beta_{ij}$ (isotropic case)  & 1.0   \\
\noalign{\smallskip}\hline
\end{tabular}
\end{table}
As a result, the generalised Hook law for a two-wave structure has the form
\begin{equation}
\label{Luk7_6}\displaystyle
\sigma^{H}_{ij}=\frac{\rho}{\rho^{*}_{0}}\textbf{C}^{*}_{ijkl}
\varepsilon^{H}_{kl}, \
\tilde{S}^{H}_{ij}=\sigma^{H}_{ij}-\alpha^{*}_{ij}
\frac{\sigma^{H}_{kl}\alpha^{*}_{kl}}{\left\|{\alpha^{*}}\right\|^{2}},
\end{equation}
where $\tilde{S}^{H}_{ij}$ is the generalised deviator of the
Hencky stress tensor. Using experimental data for through thickness
orientation, the following algorithm is developed for an accurate
extrapolation of experimental (through thickness orientation)
thermodynamic states, $\left(\sigma^{H}_{ij}\right)^{p}$, to
high-pressure shock Hugoniot states, $\left(P_{H}\right)^{p}$:
\begin{equation}
\label{Luk7_7}
\begin{array}{*{20}c}
\displaystyle
\left(\sigma^{H}_{ij}\right)^{p}=\frac{\left(\rho\right)^{p}}{\rho^{*}_{0}}
\textbf{C}^{*}_{ij33}\left(\varepsilon^{H}_{33}\right)^{p},
\hfill \hfill \hfill \hfill \hfill \hfill \\
\displaystyle
\left(\tilde{S}^{H}_{ij}\right)^{p}=\left(\sigma^{H}_{ij}\right)^{p}-\alpha^{*}_{ij}
\frac{\left(\sigma^{H}_{kl}\right)^{p}\alpha^{*}_{kl}}{\left\|{\alpha^{*}}\right\|^{2}},
\hfill \hfill \hfill \hfill \hfill \hfill \\
\displaystyle
\left(p^{s}\right)^{p}=\frac{\beta^{*}_{ij}
\left(\tilde{S}^{H}_{ij}\right)^{p}}{\beta^{*}_{kl}\alpha^{*}_{kl}},
\hfill \hfill \hfill \hfill \hfill \hfill \\
\displaystyle
\left(p^{*}\right)^{p}=-\left[\left(\sigma^{H}_{33}\right)^{p}_{\mbox{exp}}
-\left(\tilde{S}^{H}_{33}\right)^{p}\right]/\alpha^{*}_{33},
\hfill \hfill \hfill \hfill \hfill \hfill \\
\displaystyle
\left(p^{EOS}\right)^{p}=\left(p^{*}\right)^{p}-\left(p^{s}\right)^{p},
\hfill \hfill \hfill \hfill \hfill \hfill \\
\displaystyle
\left(e\right)^{p} = \frac{1}{2\rho}\left(\sigma^{H}_{33}\right)^{p}_{\mbox{exp}}
\left(\varepsilon^{H}_{33}\right)^{p},
\hfill \hfill \hfill \hfill \hfill \hfill \\
\displaystyle
\left(P^{A}_{H}\right)^{p}=
\frac{\displaystyle\left[\left(p^{EOS}\right)^{p}-\left(\rho\right)^{p}
\left(\Gamma\left(\nu\right)\right)^{p}
\cdot \left(e\right)^{p}\right]}{\displaystyle\left({1-\frac{{\left(\Gamma
\left(\nu\right)\right)^{p}
}}{{2}}\left(\mu\right)^{p}}\right)},
\hfill \hfill \hfill \hfill \hfill \hfill \\
\displaystyle
\left(U^{A}_{s}\right)^{p}=\frac{\left(P^{A}_{H}\right)^{p}}{\rho^{*}_{0}
\left(u_{p}\right)^{p}},
\hfill \hfill \hfill \hfill \hfill \hfill \\
\end{array}
\end{equation}
where the notation $\left(\bullet\right)^{p}$ denotes
interpolation point (or experimental point) $p$,
$\left(\sigma^{HSL}_{33}\right)_{\mbox{exp}}$ represents the
experimentally measured Hugoniot Stress Levels (HSLs) behind the
longitudinal (through thickness orientation) shock wave. The data
for the bulk shock wave propagation $U^{A}_{s}$ (for a single wave 
and two-wave structures) in the selected CFC material has been 
obtained and subsequently fitted by straight
lines, that is, using linear relationships of the form 
(\ref{Luk_Shock_Waves_Ani1}) and (\ref{Luk_Shock_Waves_Ani2})
(see Figures (\ref{fig:3a}), (\ref{fig:4a})).\\
For a single wave structure:
\begin{equation}
\label{Luk_Shock_Waves_Ani1}
\begin{array}{*{20}c}
\displaystyle
U^{A}_{s}=c^{A}_{0}+S^{A}_{1}u_{p}, \ c^{A}_{0}, \ S^{A}_{1}>0, \ 
\forall \ u_{p}.
\end{array}
\end{equation}
For a two-wave structure:
\begin{equation}
\label{Luk_Shock_Waves_Ani2}
\displaystyle
U^{A}_{s}=\left\{
\begin{array}{*{20}c}
\displaystyle
c^{AA}_{0}+S^{AA}_{1}u_{p}, c^{AA}_{0}, S^{AA}_{1}>0,  u_{p}\leq u^{*}_{p},
\hfill\\
c^{IA}_{0}+S^{IA}_{1}u_{p}, c^{IA}_{0}, S^{IA}_{1}>0, \ \ u_{p} > u^{*}_{p}.
\end{array}\right.
\end{equation}
The EOS data for the selected CFC material is presented in Table \ref{tab:3}. 
It is important to point out that isotropic (damaged composite) has ambient 
released density $\rho^{I}=1.40 \ g/cm^{3}$, which is less than the original 
density $\rho^{A}=1.50 \ g/cm^{3}$. This fact is explained by the damage 
softening process. 
%
%
%
%
\begin{table}
\caption{EOS data for CFC used in analysis.}
\label{tab:3}       
\begin{tabular}{lll}
\hline\noalign{\smallskip}
Parameter & Description & CFC  \\
\noalign{\smallskip}\hline\noalign{\smallskip}
$c^{A}_{0} \ \left[m/s\right]$ & Velocity curve intercept & 3590.6 \\
$S^{A}_{1}$ & First slope coefficient & 10.755 \\
$c^{AA}_{0} \ \left[m/s\right]$ & Velocity curve intercept & 3590.6 \\
$S^{AA}_{1}$ & First slope coefficient & 10.755 \\
$c^{IA}_{0} \ \left[m/s\right]$ & Velocity curve intercept & 2745.7 \\
$S^{IA}_{1}$ & First slope coefficient & 2.9119 \\
$\gamma^{A}_{0}$ & Gr\"{u}neisen gamma & 0.8500  \\
$a^{A}$ & First-order volume correction & 0.5000  \\
$c_{I} \ \left[m/s\right]$ & First anisotropic sound speed & 3599.6 \\
$c_{II} \ \left[m/s\right]$ & Second anisotropic sound speed & 2264.2 \\
$c^{L}_{0} \ \left[m/s\right]$ & Velocity curve intercept & 3228.5 \\
$S^{L}_{1}$ & First slope coefficient & 0.9203 \\
$c^{AL}_{0} \ \left[m/s\right]$ & Velocity curve intercept & 3274.0 \\
$S^{AL}_{1}$ & First slope coefficient & 1.200 \\
$c^{IL}_{0} \ \left[m/s\right]$ & Velocity curve intercept & 3145.2 \\
$S^{IL}_{1}$ & First slope coefficient & 1.0544 \\
$c^{F}_{0} \ \left[m/s\right]$ & Velocity curve intercept & 3567.7 \\
$S^{F}_{1}$ & First slope coefficient & 0.5398 \\
$c^{AF}_{0} \ \left[m/s\right]$ & Velocity curve intercept & 3933.1 \\
$S^{AF}_{1}$ & First slope coefficient & 1.2270 \\
$c^{IF}_{0} \ \left[m/s\right]$ & Velocity curve intercept & 3273.8 \\
$S^{IF}_{1}$ & First slope coefficient & 0.9405 \\
$u^{*}_{l}\ \left[m/s\right]$  & Transition particle velocity & 179.5 \\
$u^{*}_{f}\ \left[m/s\right]$  & Transition particle velocity & 333.0 \\
\noalign{\smallskip}\hline
\end{tabular}
\end{table}

Finally, having obtained all of the EOS data in terms of the shock and particle velocities, 
the following algorithm is used to obtain an accurate extrapolation of the high-pressure shock
Hugoniot states to other thermodynamic states (HSLs) for the selected
shocked carbon-fibre epoxy composite (CFC):
\begin{equation}
\label{Luk7_9}
\begin{array}{*{20}c}
\displaystyle
\left(\sigma^{H}_{ij}\right)^{T}=\frac{\rho}{\rho^{*}_{0}}\textbf{C}^{*}_{ijkl}
\varepsilon^{H}_{kl},
\hfill \hfill \hfill \hfill \hfill \hfill \\
\displaystyle
\tilde{S}^{H}_{ij}=\left(\sigma^{H}_{ij}\right)^{T}-\alpha^{*}_{ij}
\frac{\left(\sigma^{H}_{kl}\right)^{T}\alpha^{*}_{kl}}{\left\|{\alpha^{*}}\right\|^{2}},
\hfill \hfill \hfill \hfill \hfill \hfill \\
\displaystyle
U^{A}_{s}=c^{A}_{0}+S^{A}_{1}u_{p}, \ \ P^{A}_{H}=\rho^{*}_{0}U^{A}_{s}u_{p},
\hfill \hfill \hfill \hfill \hfill \hfill \\
\end{array}
\end{equation}
and
\begin{equation}
\label{Luk7_10}
\begin{array}{*{20}c}
\displaystyle
\left(p^{EOS}\right)_{i+1}=P^{A}_{H}\cdot\left({1-\frac{{\Gamma\left(\nu\right)
}}{{2}}\mu}\right)+\rho\Gamma\left(\nu\right) \left(e\right)_{i},
\hfill \hfill \hfill \hfill \hfill \hfill \\
\displaystyle
\left(p^{*}\right)_{i+1}=\left(p^{EOS}\right)_{i+1}
+\frac{\beta^{*}_{mn}\tilde{S}^{H}_{mn}}{\beta^{*}_{kl}\alpha^{*}_{kl}},
\hfill \hfill \hfill \hfill \hfill \hfill \\
\displaystyle
\left(\sigma^{HSL}_{ij}\right)_{i+1} = -\left(p^{*}\right)_{i+1}\alpha^{*}_{ij}
+\tilde{S}^{H}_{ij},
\hfill \hfill \hfill \hfill \hfill \hfill \\
\displaystyle
\left(e\right)_{i+1}=\frac{1}{2\rho}\left(\sigma^{HSL}_{ij}\right)_{i+1}
\varepsilon^{H}_{ij},
\hfill \hfill \hfill \hfill \hfill \hfill \\
\end{array}
\end{equation}
where $\left(\sigma^{H}_{ij}\right)^{T}$ represents the trial Hugoniot
Stress Levels (HSLs) behind the shock wave, $\sigma^{HSL}_{ij}$ represents
the true Hugoniot Stress Levels (HSLs) behind the shock wave,
$\rho$ is the density obtained from (\ref{Luk7_3}), $\rho^{*}_{0}$ is
the initial density (either $\rho^{A}_{0}$ or $\rho^{I}_{0}$), $\mu$ is 
the relative change of volume calculated according to (\ref{Luk4_3}) and 
$\Gamma\left(\nu\right)$ is the Gr\"{u}neisen gamma calculated according to 
(\ref{Luk4_3}). The iterative algorithm (\ref{Luk7_10}) is performed until the
following convergence criterion is achieved:
$\left|\left(e\right)_{i+1}-\left(e\right)_{i}\right| <
\left|\left(e\right)_{i}\right|\cdot error$, where
the notation $\left(\cdot\right)_{i}$ and $\left(\cdot\right)_{i+1}$
represents, respectively, physical quantities in (\ref{Luk7_9}) and
(\ref{Luk7_10}) at the $i$ and $(i+1)$ iterative steps, meanwhile
$error=10^{-5}$ represents the numerical error of the iterative algorithm.

\begin{figure}
\includegraphics[width=8.5cm]{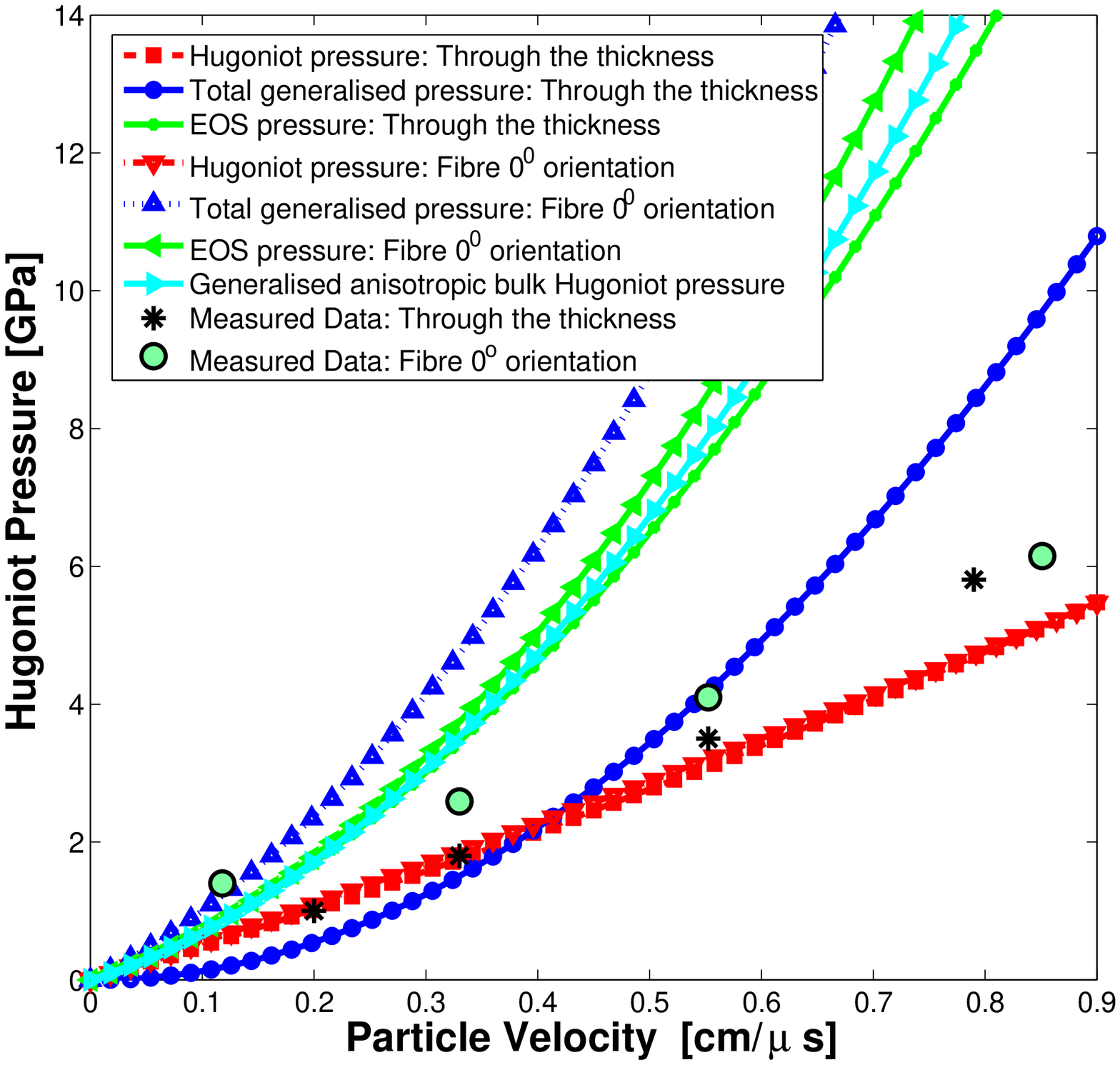}
\vspace{0.1cm}       
\caption{Comparison of the different 
Shock Hugoniot pressures $P^{L}_{H}$, $P^{*}_{L}$, $P^{EOS}_{L}$, 
$P^{F}_{H}$, $P^{*}_{F}$, $P^{EOS}_{F}$ and $P^{A}_{H}$ for a 
single-wave structure of the carbon-fibre epoxy composite behind the shock
wave in stress-particle velocity space. The $P^{L}_{H}$, $P^{F}_{H}$ and
$P^{A}_{H}$ curves are calculated from the equations
$P^{L}_{H}=\rho^{A}_{0}U^{L}_{s}u_{p}$, $P^{F}_{H}=\rho^{A}_{0}U^{F}_{s}u_{p}$ 
and $P^{A}_{H}=\rho^{A}_{0}U^{A}_{s}u_{p}$ respectively, where $U^{A}_{s}$ 
is defined by (\ref{Luk_Shock_Waves_Ani1}). The $P^{*}_{L}$, $P^{EOS}_{L}$,
$P^{*}_{F}$ and $P^{EOS}_{F}$ curves are defined by  
(\ref{Luk4_1}) and (\ref{Luk4_1_1}).} 
\label{fig:3b}       
\end{figure}

\begin{figure}
\includegraphics[width=8.5cm]{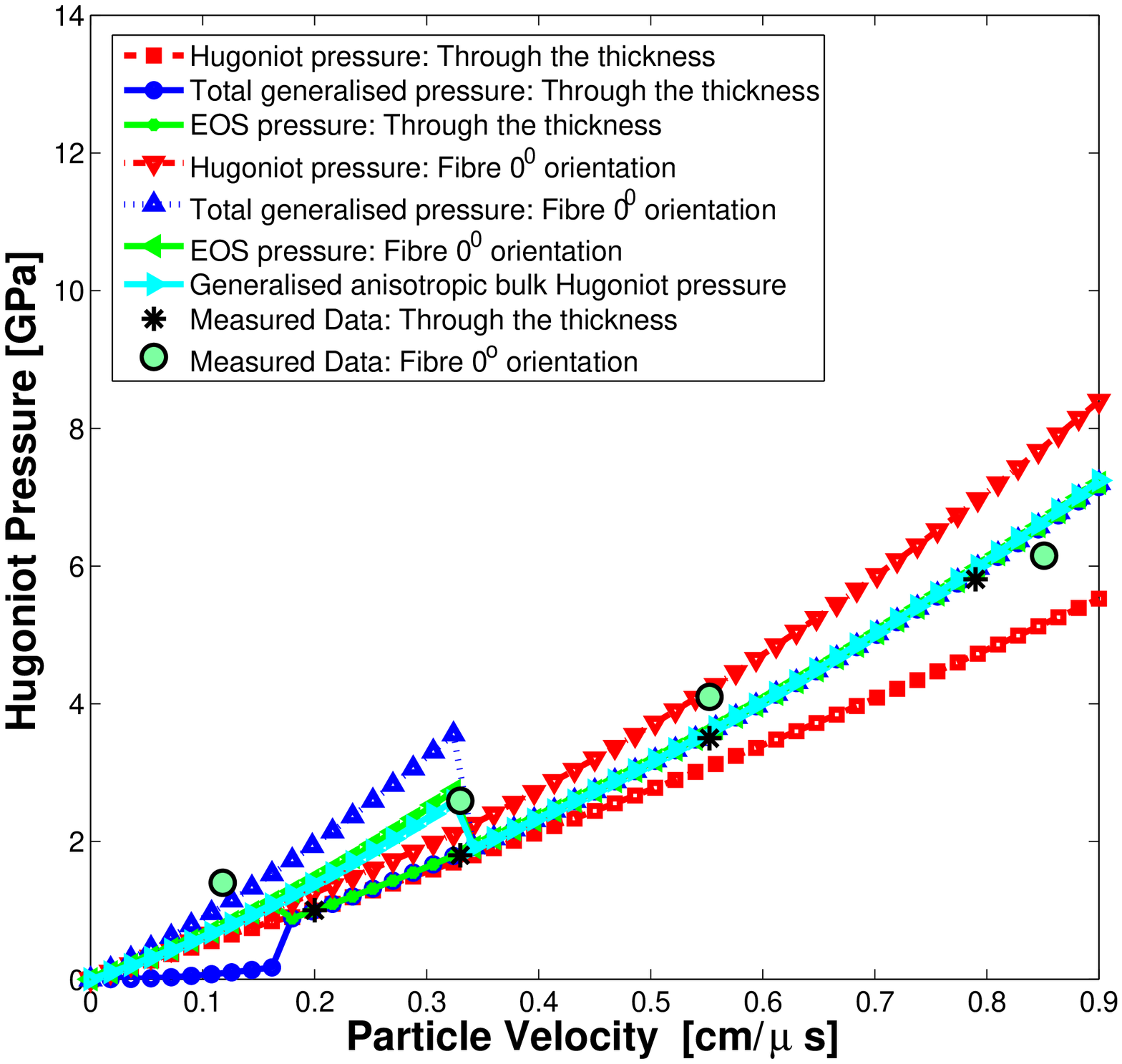}
\caption{Comparison of the different 
Shock Hugoniot pressures $P^{L}_{H}$, $P^{*}_{L}$, $P^{EOS}_{L}$, $P^{F}_{H}$, 
$P^{*}_{F}$, $P^{EOS}_{F}$ and $P^{A}_{H}$ for a two-wave 
structure of the carbon-fibre epoxy composite behind the shock
wave in stress-particle velocity space. The $P^{L}_{H}$, $P^{F}_{H}$ and
$P^{A}_{H}$ curves are calculated from equations
$P^{L}_{H}=\rho^{*}_{0}U^{L}_{s}u_{p}$, $P^{F}_{H}=\rho^{*}_{0}U^{F}_{s}u_{p}$ 
and $P^{A}_{H}=\rho^{*}_{0}U^{A}_{s}u_{p}$ respectively, where $U^{A}_{s}$ is 
defined by (\ref{Luk_Shock_Waves_Ani2}). The $P^{*}_{L}$, $P^{EOS}_{L}$,
$P^{*}_{F}$ and $P^{EOS}_{F}$ curves are defined by 
(\ref{Luk4_1}) and (\ref{Luk4_1_1}).}
\label{fig:4b}       
\end{figure}

In the following figures, the proposed anisotropic equation of
state (EOS) for the selected carbon-fibre epoxy composite (for a single 
wave and two-wave structures) is examined in terms of the (i) shock and 
particle velocities and (ii) stress (or pressure) and particle 
velocities. Figures (\ref{fig:3a}) and (\ref{fig:4a}) 
display, for a single wave and two-wave structures, respectively, the 
relationships between shock velocities and particle velocities 
for (i) through thickness and (ii) along the fibre $0^{o}$ orientation. These 
figures also show the relationship between the anisotropic generalised bulk shock 
wave and the particle velocities. Figures (\ref{fig:3b}) and 
(\ref{fig:4b}) depict, the for a single wave and two-wave structures, 
respectively, the shock Hugoniot states for (i) through thickness $P^{L}_{H}$ 
(Hugoniot pressure), $P^{*}_{L}$ (total generalised pressure), 
$P^{EOS}_{L}$ (EOS pressure) (ii) along the fibre $0^{o}$ orientation Hugoniot 
states $P^{F}_{H}$ (Hugoniot pressure), $P^{*}_{F}$ (total generalised pressure), 
$P^{EOS}_{F}$ (EOS pressure) and, also, the anisotropic generalised bulk shock 
Hugoniot states, $P^{A}_{H}$. From Fig. (\ref{fig:3a}), it follows 
that the anisotropic generalised bulk shock wave (for a single wave structure) 
has a higher velocity of propagation, for the selected carbon-fibre epoxy composite, 
compared with the longitudinal (through the thickness) shock wave, for 
all experimental and fitted points. The EOS value of $c^{A}_{0}$ in 
(\ref{Luk_Shock_Waves_Ani1}) was determined to be $3590.6\ m/s$ (for 
a single wave structure). However, Fig. (\ref{fig:4a}) shows 
different CFC behaviour for a two-wave shock structure, where the anisotropic 
generalised bulk shock wave has a lower velocity at low particle velocity 
and a higher velocity at high particle velocity. The EOS values (for a two-wave 
structure) of the velocities $c^{AA}_{0}$ and $c^{IA}_{0}$ in 
(\ref{Luk_Shock_Waves_Ani2}) were determined to be $3590.6\ m/s$ and 
$2745.7\ m/s$, respectively. In isotropic metals, the empirically 
derived EOS value of $c_{0}$ equates with the theoretical bulk sound speed. 
Note that, for a two-wave structure, the fitted EOS value of $c^{IA}_{0}=2745.7\ m/s$ 
in the isotropic region equates with the theoretical bulk sound speed $c_{B}=2745.6\ m/s$, 
and is lower than the measured anisotropic longitudinal (through the thickness) sound 
speed of $3020 \ m/s$. For the anisotropic region, the values of $c^{A}_{0}$ and 
$c^{AA}_{0}$ given above are significantly greater than the measured longitudinal (through 
the thickness) sound speed of $3020 \ m/s$. These values of $c^{A}_{0}$ and 
$c^{AA}_{0}$ are also greater than the fitted EOS values of $c^{L}_{0}=3228.5 \ m/s$ and 
$c^{F}_{0}=3567.7 \ m/s$ (for a single wave structure) and $c^{AL}_{0}=3274.0 \ m/s$ (for a 
two-wave structure) - this applies to the longitudinal through thickness and along the 
fibre $0^{o}$ orientations. However, $c^{A}_{0}$ and $c^{AA}_{0}$ are smaller than the 
fitted longitudinal (along the fibre $0^{o}$ orientation) EOS value of 
$c^{AF}_{0}=3933.1 \ m/s$ (for a two-wave structure). Note that the longitudinal (through 
thickness) fitted EOS value of $c^{L}_{0}=3230 \ m/s$ is greater than the measured 
longitudinal (through thickness) sound speed of $3020 \ m/s$ \cite{MillettLukyanov}. This 
is a behaviour that has been observed in many 
polymers, including epoxy resins \cite{Millettetal}, \cite{CarterMarsh}. Hence, for a two-wave 
structure containing damage softening effects, similar conclusions can be observed in many 
anisotropic polymers, that is, the longitudinal (through thickness) EOS value 
of $c^{AL}_{0}$ will be greater than the measured longitudinal (through thickness) sound speed. 
The longitudinal (along the fibre $0^{o}$ orientation) fitted EOS values of $c^{F}_{0}=3567.7 \ m/s$ 
(for a single wave structure) and $c^{AF}_{0}=3933.1 \ m/s$ (for a two-wave 
structure) are smaller than the respective calculated longitudinal (along the fibre $0^{o}$, 
$90^{o}$ orientations) sound speeds of $6762.0 \ m/s$ and $6666.0 \ m/s$. Furthermore, the 
fitted EOS values of $c^{A}_{0}=c^{AA}_{0}=3590.6 \ m/s$ may be compared with the first and 
second generalised anisotropic bulk speeds of sound, as the generalisation of an isotropic 
case \cite{Lukyanov3}:
\begin{equation}
\label{Luk7_12} \displaystyle
c^{A}_{0},c^{AA}_{0} \in [c_{II},c_{I}], \ c_{I}=\sqrt\frac{K_{C}}{\rho_{0}},
\ c_{II}=\sqrt\frac{K_{S}}{\rho_{0}}.
\end{equation}
Using the CFC elastic properties presented in Table \ref{tab:1}, it
can be seen that the fitted EOS value of $c^{A}_{0}=c^{AA}_{0}= 3590.6 \ m/s$ 
is in the range $[c_{II},c_{I}]$, where the analytical calculations give  
$c_{II}=2264.2 \ m/s$ and $c_{I}=3599.6 \ m/s$. It is important to re-iterate 
that, in the isotropic region, $c_{II}=c_{I}=2745.6 \ m/s$ meanwhile the fitted EOS 
value of $c^{IA}_{0}=2745.7\ m/s$ was obtained.
 
The experimental shock velocity in the fibre $0^{o}$ orientation is initially 
greater than that corresponding to the through thickness orientation. In time, the 
shock velocity decreases with pressure and, eventually, there is convergence 
between these data sets. A number of mechanisms have been proposed to explain 
this behaviour \cite{Bordzilovskyetal,Hereiletal,MillettLukyanov}. However, 
the experimental data shows that, at lower particle velocities, the stress pulse 
is transmitted through an anisotropic mixture of epoxy binder and fibres (see 
Table \ref{tab:1}), whereas at higher particle velocities, this pulse is 
transmitted through an isotropic mixture instead (see Table \ref{tab:2}). The stable 
shock waves (where shock velocity decreases with pressure) can exist when the shock 
front breaks up into two or more waves \cite{Bethe}. As is shown above, the two-wave 
front structure is sufficient to fit experimental data for the orientations through the 
thickness and along the fibre $0^{o}$. 
    
Using a generalised decomposition of the stress tensor, the generalised anisotropic 
bulk Hugoniot pressure ($P^{A}_{H}=\rho^{*}_{0}U^{A}_{s}u_{p}$) can be defined and 
compared to the longitudinal Hugoniot pressures ($P^{L}_{H}=\rho^{*}_{0}U^{L}_{s}u_{p}$ 
for through the thickness; $P^{F}_{H}=\rho^{*}_{0}U^{F}_{s}u_{p}$ for along the fibre 
$0^{o}$ orientation), as shown in Fig. (\ref{fig:3b}) for a single wave 
structure and in Fig. (\ref{fig:4b}) for a two-wave structure. 
It can be seen that, for a single wave structure, there is a significant difference 
between the longitudinal Hugoniot pressures, $P^{L}_{H}$ and $P^{F}_{H}$, and the calculated 
generalised anisotropic bulk Hugoniot pressure, $P^{A}_{H}$. This indicates that, for 
highly anisotropic materials and the assumption of a single wave structure, the anisotropic 
bulk shock front will be supersonic with respect to the longitudinal shock front in 
the least stiff direction (e.g., the through thickness orientation). However, this is not the 
case in the two-wave structure. 
%
%
%
%
%
%
%
%
%
%
%
%
%
According to the generalised decomposition (\ref{Luk3_1}), the 
stress tensor has been split into the generalised spherical component, 
$p^{\ast}\alpha_{ij}$, and the generalised 
deviatoric component, $\tilde{S}_{ij}$. The total generalised anisotropic  
"pressure", $p^{\ast}$, comprises a sum of two terms, 
$p^{*}=p^{EOS}+p^{\tilde{S}}$, where $p^{EOS}$ corresponds to the thermodynamic 
(EOS) response calculated from (\ref{Luk4_2}) - (\ref{Luk4_3}) and $p^{\tilde{S}}$
corresponds to the generalised deviatoric stress calculated using (\ref{Luk3_4}).

Figures (\ref{fig:3b}) and (\ref{fig:4b}) display $p^{*}$ and 
$p^{EOS}$. It can be seen that there is an increasing divergence of $p^{\ast}$ from 
$p^{EOS}$ due to greater particle velocities associated with an increased contribution 
from $p^{\tilde{S}}$. In addition, it can be seen that pressure $p^{EOS}$ for the selected 
CFC material is greater than the generalised total pressure, $p^{\ast}$. Further comparison 
of different pressures (see Figs. (\ref{fig:3b}) and 
(\ref{fig:4b})) shows that there is an increasing divergence 
between the measured Hugoniot stress (HSLs) and the calculated pressures 
$P^{F}_{H}$, $P^{L}_{H}$, $p^{EOS}_{F}$, $p^{EOS}_{L}$, $p^{*}_{F}$ and  
$p^{*}_{L}$ - this corresponds to the increasing 
contribution from the generalised deviator of the stress tensor, $\tilde{S}^{H}_{ij}$.
In addition, it is important to analyse the difference between the shock velocities 
in the two-wave structure at the phase transition point, these evaluated using:
\begin{equation}
\label{Luk7_13} \displaystyle
\left[U^{L}_{s}\right]=\left(c^{AL}_{0}-c^{IL}_{0}\right)
+\left(S^{AL}_{1}-S^{IL}_{1}\right)u_{p},
\end{equation}
\begin{equation}
\label{Luk7_13_1} \displaystyle
\left[U^{F}_{s}\right]=\left(c^{AF}_{0}-c^{IF}_{0}\right)
+\left(S^{AF}_{1}-S^{IF}_{1}\right)u_{p},
\end{equation}
where $\left[U^{L}_{s}\right]_{u_{p}=u^{*}_{l}}=154.935\ m/s$ for the 
through thickness orientation and 
$\left[U^{F}_{s}\right]_{u_{p}=u^{*}_{f}}=754.705\ m/s$ for the 
fibre $0^{o}$ orientation at the transition points $u^{*}_{l}$ 
and $u^{*}_{f}$ respectively. Using fitted EOS data (see Table 
\ref{tab:3}), the difference between shock waves at the transition point 
for the through thickness orientation is smaller than that along the 
fibre $0^{o}$ orientation. Therefore, the material behaviour through the 
thickness orientation can be approximated with a single wave structure 
(see Figure \ref{fig:3a}), this confirmed during the 
examination of Hugoniot Stress Levels (HSLs) through the 
thickness orientation in terms of the particle velocities \cite{Lukyanov7}.

\begin{figure}
\includegraphics[width=8.5cm]{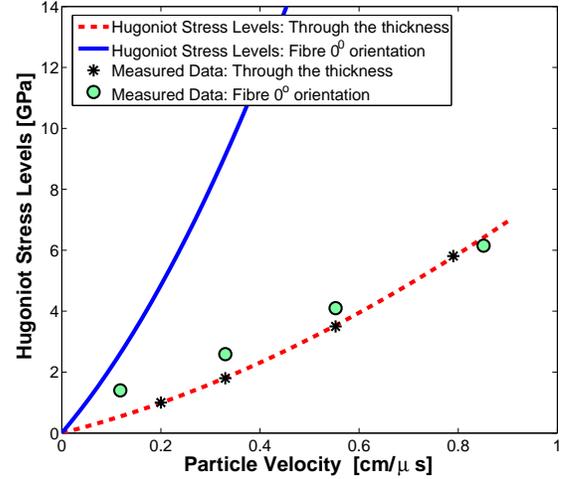}
\caption{Hugoniot Stress Levels (HSLs) for a carbon-fibre 
epoxy composite behind the shock wave in the through thickness and fibre 
$0^{o}$ orientations in stress-particle velocity space, assuming a single 
wave structure. The experimental data for Hugoniot Stress Levels (HSLs) 
was obtained by Millett {\it et al.} \cite{MillettLukyanov}. The dotted 
curve is calculated using the anisotropic EOS proposed in (\ref{Luk7_9}), 
(\ref{Luk7_10}) and experimental data for the through thickness orientation 
$U^{L}_{S}-u_{p}$. The solid curve is calculated using the anisotropic EOS 
proposed in (\ref{Luk7_9}), (\ref{Luk7_10}) and experimental data for the 
fibre $0^{o}$ orientation $U^{F}_{S}-u_{p}$.} 
\label{fig:5a}       
\end{figure}

\begin{figure}
\includegraphics[width=8.5cm]{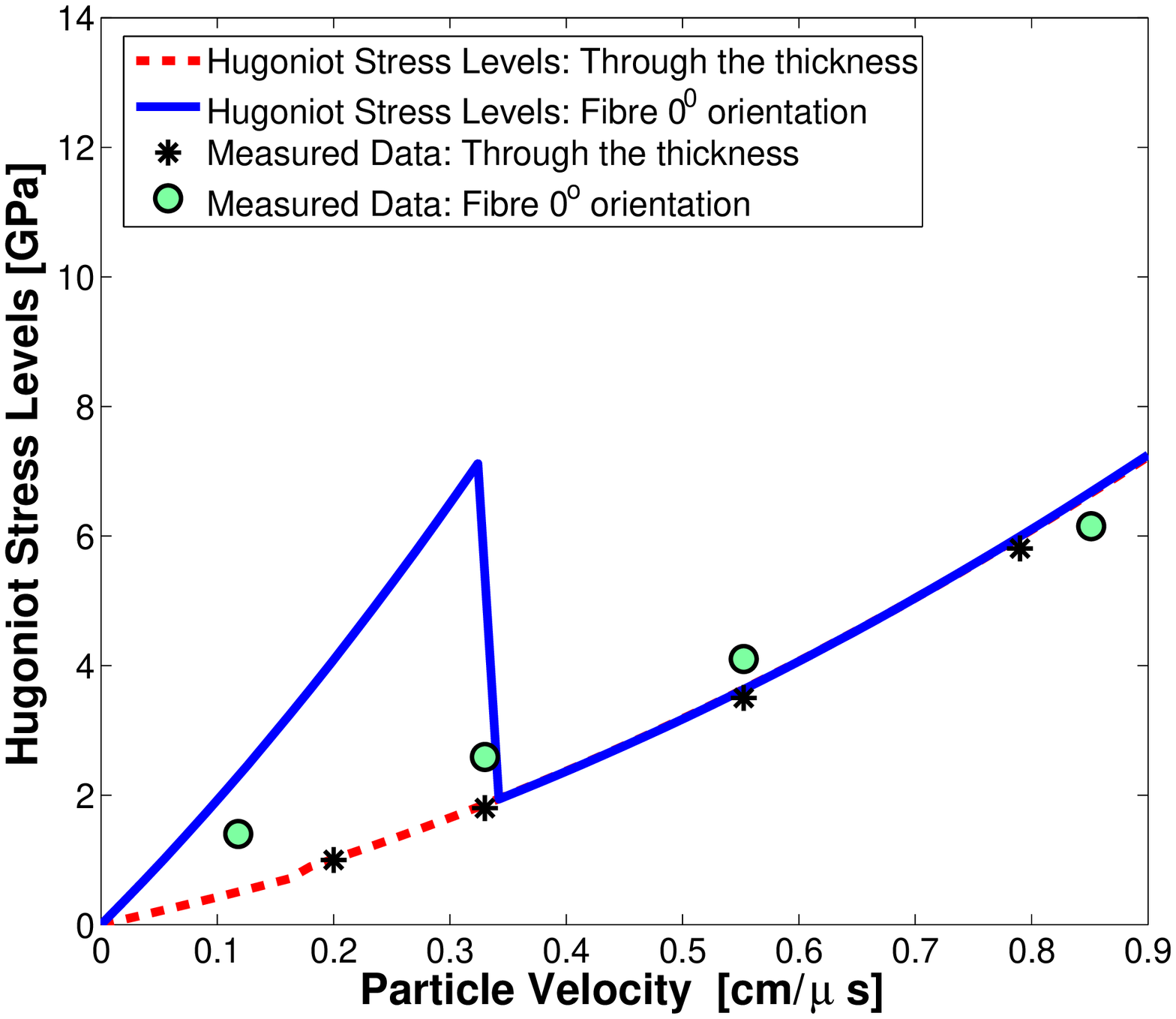}
\caption{Hugoniot Stress Levels (HSLs) 
for a carbon-fibre epoxy composite behind the shock wave in the through 
thickness and fibre $0^{o}$ orientations in stress-particle velocity 
space, assuming a two-wave structure. The experimental data for 
Hugoniot Stress Levels (HSLs) was obtained by Millett {\it et al.} 
\cite{MillettLukyanov}. The dotted curve is calculated using the 
anisotropic EOS proposed in (\ref{Luk7_9}), (\ref{Luk7_10}) and experimental 
data for the through thickness orientation $U^{L}_{S}-u_{p}$. The solid curve 
is calculated using the anisotropic EOS proposed in (\ref{Luk7_9}), (\ref{Luk7_10}) 
and experimental data for the fibre $0^{o}$ orientation $U^{F}_{S}-u_{p}$.}
\label{fig:5b}       
\end{figure}

Finally, in Figs. (\ref{fig:5a}) and (\ref{fig:5b}), 
the effect of orientation on the Hugoniot Stress Levels (HSLs) in 
stress-particle velocity space is examined for single wave and two-wave
structures. It can be seen from the experimental points in Figs. 
(\ref{fig:5a}), (\ref{fig:5b}) that, at lower stresses, the mixture of 
epoxy binder and $0^{o}$ fibres is stiffer, however, as stress increases, 
the experimental Hugoniots for a mixture (CFC composite) of both orientations 
converge. This is in agreement with the behaviour of the shock velocities shown 
in Fig. (\ref{fig:5b}) for a two-wave structure. It is also
clear from Fig. (\ref{fig:5a}) that a single wave structure
is not capable of predicting correctly the Hugoniot Stress Levels (HSLs) 
along the fibre $0^{o}$ orientation to agree with the stability requirements 
formulated by Bethe \cite{Bethe}. It is important to note that 
there is no true prediction of the experimental data shown in figures 
(\ref{fig:5a}) and (\ref{fig:5b}) for through the 
thickness orientation. These experimental points (through the thickness) 
have been used to define the material parameters in the presented anisotropic 
EOS model. There is only a true prediction of the experimental data 
shown in figures (\ref{fig:5a}) and (\ref{fig:5b}) for 
along the fibre $0^{o}$ orientation.     
%
%
%
%
%
%
Figures (\ref{fig:5a}), (\ref{fig:5b}) show qualitatively that the anisotropy 
of a composite material (carbon fibre-epoxy composite) has a strong effect 
on the accurate extrapolation of high-pressure shock Hugoniot states to other 
thermodynamic states for shocked anisotropic composite (CFC) materials (e.g., a
carbon-fibre epoxy composite) of any symmetry.
\section{Conclusions}\label{sec:9}
An anisotropic equation of state is proposed for the accurate 
extrapolation of high-pressure shock Hugoniot states to other 
thermodynamic states, for a shocked carbon-fibre epoxy composite 
(CFC) of any symmetry. The proposed equation of state, which uses 
a generalised decomposition of the stress tensor 
\cite{Lukyanov2,Lukyanov3,Lukyanov4,Lukyanov1,Lukyanov5}, 
represents the mathematical and physical generalisation of the 
Mie-Gr\"{u}neisen equation of state for an isotropic material, and 
reduces to the latter in the limit of isotropy.

Further insights into the anisotropic CFC response under shock 
loading can be gained from an examination of the material 
EOS in terms of: shock stresses (total generalised anisotropic 
pressure, $p^{*}$; generalised anisotropic bulk Hugoniot pressure, 
$P^{A}_{H}$; pressure, $p^{EOS}$, corresponding to the thermodynamic 
(equation of state) response; pressure, $p^{\tilde{S}}$, corresponding 
to the generalised deviatoric stress), shock velocities (shock 
velocity in the through-thickness orientation, $U^{L}_{S}$, shock velocity 
in the fibre $0^{o}$ orientation, $U^{F}_{S}$, and the generalised anisotropic 
bulk shock velocity, $U^{A}_{s}$) and the particle velocity, $u^{p}$. Figure
(\ref{fig:4a}) shows linear relationships, for a two-wave 
structure, between the shock velocities $U^{L}_{s},U^{F}_{S}, U^{A}_{S}$ 
and the particle velocities, $u_{p}$, over the range of measurements made 
during experiments. The values of $c^{AL}_{0}, c^{AF}_{0}, c^{AA}_{0}$ and  
$c^{IL}_{0}, c^{IF}_{0}, c^{IA}_{0}$ were determined. The values  
$c^{AA}_{0}$ and $c^{IA}_{0}$ (the intercept of the $U^{A}_{s}$-$u_{p}$ 
curve for two-wave structure) are in the interval between the first and 
second generalised anisotropic bulk speeds of sound \cite{Lukyanov3} 
(for non-linear anisotropic elastic and isotropic elastic shock waves). 
This is a behaviour that is observed in many polymers, including 
epoxy resins. When $c^{AL}_{0}$, $c^{AF}_{0}$ and $c^{AA}_{0}$ are compared to 
the measured longitudinal sound speed $C_{L}$ (Table \ref{tab:1}, 
Table \ref{tab:3}), it can be seen that the former values are significantly 
greater than the latter, in the through thickness orientation. This indicates that, 
for highly anisotropic materials, anisotropic shock fronts (at lower particle velocity) 
are always supersonic with respect to the longitudinal sound speed in the 
least stiff direction (for example, longitudinal sound speed in the through 
thickness orientation). It is possible that the generalised anisotropic 
bulk shock velocity, $U^{A}_{s}$, depends non-linearly on particle 
velocity for a given anisotropic material (or composite materials), 
as has been shown for some isotropic polymers such as PMMA 
\cite{BarkerHollenbach} -- unfortunately, the corresponding 
experimental data for carbon fibre materials (or other 
anisotropic materials) cannot be located at this time. 

An analytical calculation showed that the Hugoniot Stress Levels (HSLs) for 
a carbon-fibre epoxy composite do not agree with the experimental data 
for a single wave structure methodology for the stability requirements formulated 
by Bethe \cite{Bethe}. However, the material behaviour in the through-thickness 
orientation can be approximated by a single wave structure due to the lack of any 
significant discontinuity at the transition zone \cite{Lukyanov7}. This approximation 
will not take into account changes in the material elastic properties during the damage 
softening process. In addition, an analytical calculation showed that the Hugoniot Stress 
Levels (HSLs) in different directions, for a CFC composite subject to the two-wave 
structure (non-linear anisotropic and isotropic elastic waves), agree with 
experimental measurements at both low shock intensities (where the $0^{0}$ 
orientation was significantly stiffer than the through-thickness orientation) 
and at high shock intensities (where the HSLs of the two orientations 
converged due to the presence of damage softening), this also in 
agreement with the stability requirements formulated by Bethe \cite{Bethe}.
\section{Acknowledgments}
\label{sec:10}

Author thanks Prof. V. Penjkov, Dr. B. Cox and Dr. B. Wells for many useful 
suggestions regarding this work.  The discussions regarding the shock wave 
experiments on a carbon-fibre epoxy composite with Dr. J. C. F. Millett 
during the meetings at Cranfield University are also greatly appreciated.
%


\begin{thebibliography}{}
%
%
\bibitem[1]{BarkerHollenbach}L.M.~Barker, R.E.~Hollenbach, J. Appl. Phys. \textbf{43},
4669 (1972)
%
\bibitem[2]{Kanel}G.I.~Kanel, J. Mech. Phys. Solids \textbf{43} (10), 1869 (1998)
%
\bibitem[3]{Kaneletal}G.I.~Kanel, K.~Baumung, H.~Bluhm, V.E.~Fortov, Nucl.
Instr. Meth. Phys. Res. A \textbf{415}, 509 (1998)
%
\bibitem[4]{BourneStevens}N.K.~Bourne, G.S.~Stevens, Rev. Sci. Instrum. \textbf{72} (4),
2214 (2001)
%
\bibitem[5]{Bourne}N.K.~Bourne, Meas. Sci. \& Technol. \textbf{14}, 273 (2003)
%
\bibitem[6]{DavisonGraham} L.~Davison, R.A.~Graham,  Shock Compres. solids Phys. Rep.
\textbf{55}, 255 (1979)
%
\bibitem[7]{Bushmanetal}A.V.~Bushman, G.I.~Kanel, A.L.~Ni, V.E.~Fortov, \textit{Intense
dynamic Loading of Condensed Matter} (Taylor and Francis, Washington, D.C., 1993)
%
\bibitem[8]{Meyers}M.A.~Meyers, \textit{Dynamic Behavior of Materials} (Wiley, Inc.,
New York, 1994)
%
\bibitem[9]{Steinberg}D.J.~Steinberg, Report No. UCRL-MA-106439, Lawrence
Livermore National Laboratory, Livermore, CA (1991)
%
\bibitem[10]{KiselevLukyanov} A.B.~Kiselev, A.A.~Lukyanov,
Int. J. Forming Processes \textbf{5}, 359 (2002)
%
\bibitem[11]{Dandekaretal}D.P.~Dandekar, C.A.~Hall, L.C.~Chhabildas,
W.D.~Reinhart, Compos. Struc. \textbf{61}, 51 (2003)
%
\bibitem[12]{MunsonMay} D.E.~Munson, R.P.~May, J. Appl. Phys. \textbf{43}, 962 (1972)
%
\bibitem[13]{Millettetal} J.C.F.~Millett, N.K.~Bourne , N.R.~Barnes, J. Appl.
Phys. \textbf{92}, 6590 (2002)
%
\bibitem[14]{Zhuketal}A.Z.~Zhuk, G.I.~Kanel, A.A.~Lash, J. Phys. IV
\textbf{4}, 403 (1994)
%
\bibitem[15]{Riedeletal}W.~Riedel, H.~Nahme, K.~Thoma, In: Furnish MD,
Gupta YM, Forbes JW, editors. Shock compression of condensed matter – 2003.
Melville, N.Y.: AIP Press, 701 (2004)
%
\bibitem[16]{Zaretskyetal}E.~Zaretsky, G.~deBotton, M.~Perl, Int.
J. Solids Struct. \textbf{41}, 569 (2004)
%
\bibitem[17]{Chenetal}J.K.~Chen, A.~Allahdadi, T.~Carney,
Comp. Sci. and Techn. \textbf{57}, 1268 (1997)
%
\bibitem[18]{Hayhurstetal} C.J.~Hayhurst, S.J.~Hiermaier,
R.A.~Clegg, W.~Riedel, and M.~Lambert, Int. J.
Impact Engineering \textbf{23}(1), 365 (1999)
%
\bibitem[19]{Andersonetal1} C.E.~Anderson, Jr.P.E.~O'Donoghue, D.~Skerhut,
J. Comp. Materials \textbf{24}, 1159 (1990)
%
\bibitem[20]{Andersonetal2} C.E.~Anderson, P.A.~Cox, G.R.~Johnson, P.J.~Maudlin,
Comput. Mech. \textbf{15}, 201 (1994)
%
\bibitem[21]{Bordzilovskyetal}S.A.~Bordzilovsky, S.M.~Karakhanov,
L.A.~Merzhievsky, In: Schmidt S.C., Dandekar D.P., Forbes J.W.,
editors. Shock compression of condensed matter
– 1997. Melville, N.Y.: AIP Press, 545 (1998)
%
\bibitem[22]{Hereiletal}P-L~Hereil, O.~Allix, M.~Gratton, J.
Phys. IV \textbf{7}, 529 (1997)
%
\bibitem[23]{MillettLukyanov} J.C.F.~Millett, N.K.~Bourne, Y.J.E.~Meziere,
R.~Vignjevic, A.A.~Lukyanov, Comp. Sci. and Techn.
\textbf{67}(15-16), 3253 (2007)
%
\bibitem[24]{Lukyanov2}A.A.~Lukyanov, Int. J. Plasticity \textbf{24}, 140 (2008)
%
\bibitem[25]{Lukyanov3}A.A.~Lukyanov, Eur. Phys. J. B \textbf{64}, 159 (2008)
%
\bibitem[26]{Lukyanov4}A.A.~Lukyanov, J. Appl. Mech. \textbf{76}, 061012-1 (2009)
%
\bibitem[27]{Lukyanov1}A.A.~Lukyanov, ASME Proceeding IPC2006,
ISBN 0-7918-3788-2 (2006)
%
\bibitem[28]{Lukyanov5}A.A.~Lukyanov, J. Pressure Vessel Technology \textbf{130},
021701-1 (2008)
%
\bibitem[29]{Lukyanov6}A.A.~Lukyanov, V.B. Penjkov, J. Appl. Math. Mech. \textbf{73} (4),
(2009)
%
\bibitem[30]{Bethe}H.A.~Bethe, Office of Scientific Res. and Develop. Rept. No. 545, 
Serial No. 237 (1942).
%
\bibitem[31]{Poirier}J.P.~Poirier, \textit{Introduction to the Physics of the Earth's Interior} 
(Cambridge: Univ. Press, 2000)
%
\bibitem[32]{CarterMarsh}W.J.~Carter, S.P.~Marsh, Report No. LA-12006-MS, Los Alamos National
Laboratory, LA (1995)
\bibitem[33]{Lukyanov7}A.A.~Lukyanov, Mech. Advan. Mater. and Struct., Special Issue: ICCS15.
Accepted (2009)  
\end{thebibliography}
\end{document}